\documentclass[%
reprint,
 amsmath,amssymb,
prb
]{revtex4-2}

\usepackage{graphicx}
\usepackage{dcolumn}
\usepackage{bm}
\usepackage{xcolor}
\usepackage{placeins}
\usepackage{float}
\usepackage[english]{babel}

\begin{document}

\preprint{APS/123-QED}

\title{Spin transport analysis for a spin pseudovalve-type L\textsubscript{l}/SC/L\textsubscript{r} trilayer\\ for L = \{FeCr, Fe, Co, NiFe, Ni\} and SC = \{GaSb, InSb, InAs, GaAs, ZnSe\}}

\author{Julián A. Zúñiga}
\email{jzuniga@iflp.unlp.edu.ar}

 \author{Arles V. Gil Rebaza}%
 \affiliation{Instituto de Física La Plata IFLP-CONICET, CCT La Plata. Departamento de Física, Facultad de Ciencias Exactas, Universidad Nacional de La Plata UNLP, 1900 La Plata, Argentina}%


\author{Diego F. Coral}
\affiliation{Grupo de Ciencia y Tecnología de Materiales Cerámicos CYTEMAC, Departamento de Física, Facultad de Ciencias Naturales, Exactas y de la Educación, Universidad del Cauca UNICAUCA, 190002 Popayán, Colombia}%




\date{\today}

\begin{abstract}

In this work, we present a theoretical study of spin transport in a trilayer pseudospin-valve (PSV) heterostructure composed of electrode ($\mathrm{L}_l$)/insulator/electrode ($\mathrm{L}_r$). The insulating layer corresponds to a semiconductor (SC) with a zinc-blende crystal structure from the III–V (GaSb, InSb, InAs, and GaAs) or the II–VI (ZnSe), while the electrodes are ferromagnetic materials $\mathrm{L}_j=\{\mathrm{FeCr}, \mathrm{Fe}, \mathrm{Co}, \mathrm{NiFe}, \mathrm{Ni}\}$. This combination yields 125 possible PSV configurations. The theoretical model implemented is based on the approach proposed by \emph{J. C. Slonczewski}. In our approach, the exchange splitting in the ferromagnetic materials and the spin–orbit coupling (SOC) of the \emph{Dresselhaus} and \emph{Rashba} types in the semiconductors are included, allowing control of the wave vector associated with the spin states.

The tunnel magnetoresistance (TMR) is calculated at low temperature as a function of the semiconductor thickness, parameterized with respect to the crystallographic axis that favors the magnetization direction in the ferromagnetic electrodes, within the \emph{Landauer–Büttiker} formalism in the single-channel regime. The results show that the TMR reaches its maximum value independently of the relative orientation between the magnetization vector and the crystallographic direction. The most efficient configuration corresponds to $\mathrm{Fe}_{90}\mathrm{Cr}_{10}$/GaSb/$\mathrm{Fe}_{90}\mathrm{Cr}_{10}$, with a TMR value of 83.60\%. Furthermore, the \emph{Dresselhaus} SOC contributes more significantly to the TMR than the \emph{Rashba} SOC. Finally, the TMR varies when the electrodes $\mathrm{L}_j$ are permuted, due to differences in their Fermi energies. The obtained results are compared with previous studies reported in the literature based on alternative theoretical frameworks or assumptions, showing good agreement.
\end{abstract}
\maketitle


\section{\label{Int}Introduction}

Spin-polarized electrons can be generated in nonmagnetic (NM) materials through several mechanisms, the most widely used of which is spin injection from a ferromagnetic (FM) material \cite{Hiro14}. In this process, FM materials—such as conventional FM metals (Fe, Co, and Ni), half-metallic ferromagnets (HMFs), diluted magnetic semiconductors (DMSs), or ferromagnetic semiconductors (FMSs)—are coupled to a NM metal or a semiconductor (SC) through either an ohmic contact or a tunneling barrier, enabling the selective transfer of electrons according to their spin orientation \cite{Hiro20}. This principle constitutes the basis of several spintronic architectures. In particular, pseudospin-valve (PSV) heterostructures are characterized by an electrode/insulator/electrode configuration, where the electrodes correspond to the FM materials mentioned above, and the intermediate layer acts as an insulating barrier associated with NM or SC materials. In such structures, the magnetoresistive properties arise from spin-dependent electronic transport through the barrier, allowing the current to be controlled by the magnetic configuration of the electrodes and giving rise to characteristic phenomena such as tunnel magnetoresistance (TMR), an effect that has been extensively studied \cite{Hiro20, Gani20, Kumar22, Takase20}.

For a theoretical description of this type of spin-dependent transport, several models have been developed to explain the transmission of charge and spin currents between FM layers separated by an insulating spacer \cite{Vincent02}. In this context, the present work considers the spin-current model proposed by \emph{Slonczewski} \cite{Slonc89}, which analyzes the coupling arising from the torque produced by the rotation of the magnetization of one FM layer with respect to the other. This description takes into account the internal exchange interaction between electrons, which leads to spin polarization in the system \cite{Yang97, Zhang97}. The coupling is described in terms of the probability of a spin-flip current, calculated from the stationary wave functions obtained from the Schrödinger equation for a free electron. Within this theoretical framework, the system is modeled by considering the splitting of the electronic energies for majority and minority spins, known as the exchange splitting ($\Delta_{xs}$) \cite{Tao04, Saffar06, Bunder07}; in addition, spin–orbit coupling (SOC) is incorporated, which influences spin-dependent transport both qualitatively and quantitatively \cite{Pope05}. In particular, the \emph{Dresselhaus} and \emph{Rashba} SOC terms in zinc-blende semiconductors (SCs) are considered. Finally, the conductance is evaluated within the \emph{Landauer–Büttiker} formalism \cite{Wim09, Matos09}, from which the TMR is obtained \cite{Tao04, Bunder07, Takase20}. In this study, the formulation is applied to a single-channel system at $T \approx 0$ K, allowing the description of electronic transport in nanometric systems \cite{David97}.

Several theoretical works \cite{Peralta06, Wang08, Autes10, Shukla23}, which have not necessarily employed the same model proposed in this work, as well as experimental studies \cite{Jiang03, Zenger04, Scheike21} on PSV structures, motivated the compilation of an initial set of conventional FMs, to which the compound FeCr was added following the suggestion by Tat-Sang Choy \emph{et al}. \cite{Choy99} that report an abrupt changes in spin polarization may occur as a function of Cr concentration \cite{Choy99, Korz09}, a feature that may influence spin-injection processes, together with insulating barriers formed by zinc-blende semiconductors (SCs), such as GaSb, InSb, InAs, GaAs, and ZnSe, with the purpose of performing a comparison of the TMR among all possible PSV combinations that can be analyzed within the model (125 in total), in order to identify relationships between the \emph{Fermi} energy ($E_F$) and $\Delta_{xs}$ of the electrodes and to suggest potentially suitable configurations for experimental studies, as well as to contrast the proposed theoretical model with previously reported results in the literature.
\section{Theoretical Model}
\begin{figure}[h!]
\includegraphics[width=0.48\textwidth]{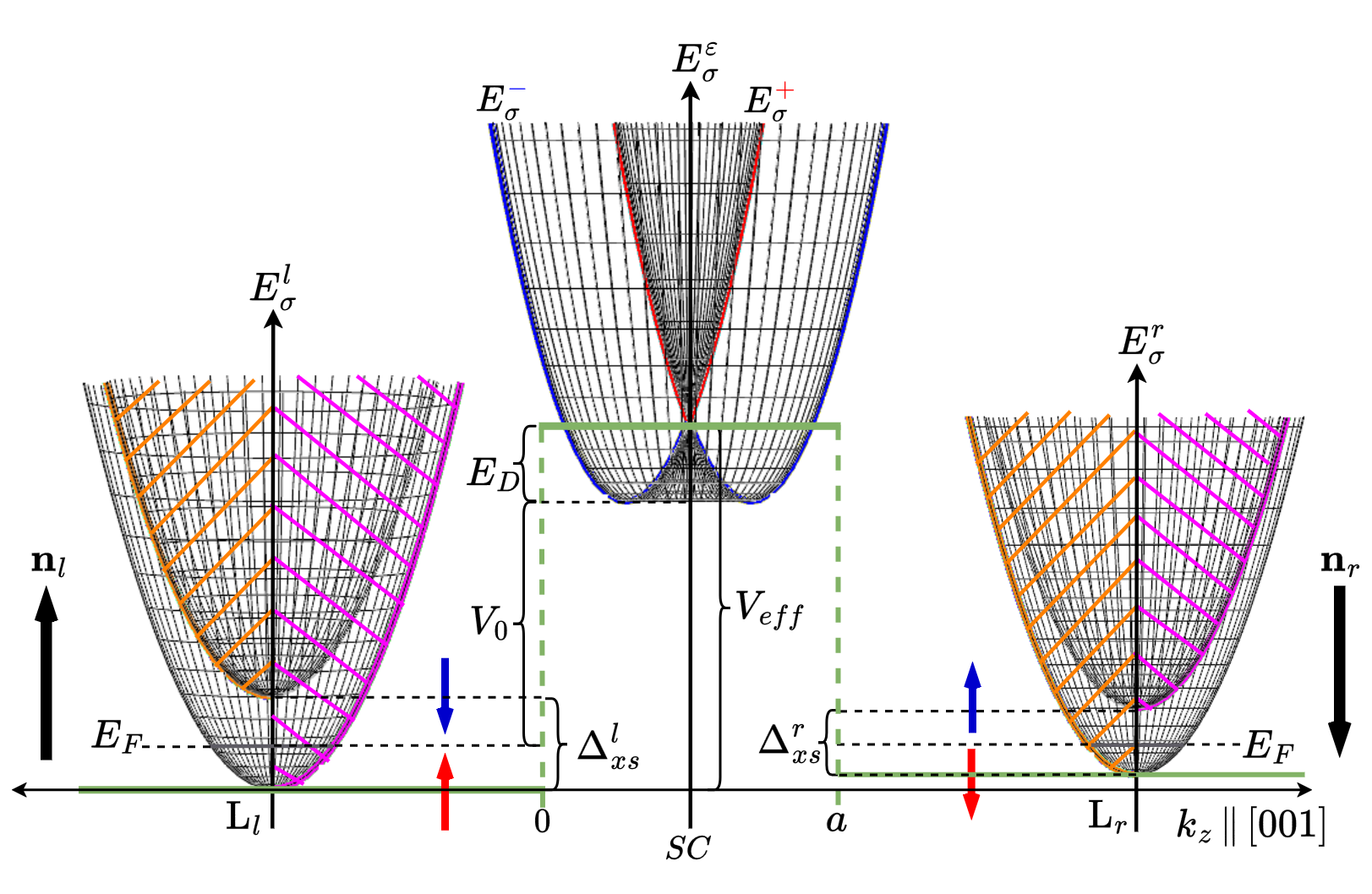}
\caption{\label{F1} Schematic representation of a rectangular potential barrier (antiparallel magnetic configuration), defined by the energy eigenvalues of the FM ($E_{\sigma}^j$) and  of the SC with SOC ($E_{\sigma}^{\varepsilon}$), along the growth direction of the heterostructure $[0\,0\,1]$, where $V_{eff}$ and $a$ denote the effective barrier height and thickness, respectively. Here, $\Delta^j_{xs}$ represents the exchange splitting, and the magnetization vector $\mathbf{n}_r$ rotates parallel to the barrier plane.}
\end{figure}

 The study of spin-dependent electronic transmission is carried out under the assumption that the potential energy profile consists of a thin rectangular barrier (1--6 nm) with an effective barrier height $V_{eff}$, defined as the sum of the $E_F$, the potential energy  $V_0$ associated with the SC band gap ($E_g$), where $E_g/2\leq V_{0} < E_g$ \cite{Zenger04}, and the \emph{Dresselhaus} SOC energy $E_D$ (see Fig. \ref{F1}). The tunneling direction is parallel to the $z$-axis, and the in-plane wave vector $\mathbf{k}_{\|j}$ (see Suppl. Mater., Sec. A, Fig. S1) is conserved throughout the heterostructure, allowing the decoupling of motion along the $z$-axis from the other spatial degrees of freedom.

\subsection{Description of the hamiltonians}

The Hamiltonian for the regions $z\leq 0$ and $z>a$, corresponding to the $\mathrm{L}_l$ and $\mathrm{L}_r$ layers, respectively, is defined as follows\cite{Bunder07, Matos09}:
\begin{equation}
  \mathcal{\hat{H}}_j=\left[\frac{\hbar^2}{2m_j^*}\left(k_{z\,j}^2+k_{\|j}^2\right)\right]\mathbb{\hat{I}} -\frac{\Delta^j_{xs}}{2}\,\mathbf{n}_j\cdot \boldsymbol{\tau},
  \label{1}
\end{equation} where $\hbar$ is the reduced {\em Planck} constant, $m_j^*$ is the effective mass in layer $\mathrm{L}_j$, and $k_{\epsilon\,j}$ (with $\epsilon = \left(x, y, z\right)$) is an operator defined as $k_{\epsilon\,j} = -i\partial_{\epsilon}$. The expression $k_{\|\,j} = (k_{x\,j},k_{y\,j})$ represents the in-plane wave vector on the barrier. $\mathbb{\hat{I}}$ corresponds the identity matrix (2$\times$2), while $\boldsymbol{\tau} = (\hat{\tau}_x, \hat{\tau}_y, \hat{\tau}_z)$ represents the vector of {\em Pauli} matrices. The magnetization normal vector in the barrier plane is given by $\mathbf{n}_j= \left(\sin\theta_j,\cos\theta_j,0\right)$. Taking the easy magnetization axis as a reference, it is convenient to express the direction of this vector relative to that axis by introducing the displacement angle \cite{Matos09}, $\phi_j = \theta_j+\theta_m$ (see Suppl. Mater., Sec. A, Fig. S2).

Therefore, we define the spinor $|\chi_\sigma^j\rangle$ that allows us to diagonalize the Hamiltonian as follows:
\begin{equation}
 |\chi_\sigma^j\rangle=\frac{1}{\sqrt{2}}\left(
     \begin{matrix}
     1 \\
     i\sigma e^{-i\theta_l}\\
     \end{matrix}
     \right),
     \label{3}
\end{equation}

Consequently, the energy eigenvalues for the layers $\mathrm{L}_j$ are described by the expression:
\begin{equation}
  E_\sigma^j=\frac{\hbar^2}{2m_j^*}k_j^2-\sigma\frac{\Delta^j_{xs}}{2}\cos\left(\delta_j\,\theta\right),
  \label{4}
\end{equation} with
$\delta_j = \left\{
       \begin{array}{lcc}
             0 & if & j = l\\
             1 & if & j = r
        \end{array},
                \right.$
where $\theta$ represents the angle between $\mathbf{n}_l$ and $\mathbf{n}_r$ (see Suppl. Mater., Sec. A).

The Hamiltonian with SOC for the region $0<z\leq a$  is given by \cite{Lu10}:
 \begin{eqnarray}
  \mathcal{\hat{H}}_{\sigma}^\varepsilon = &&\left(\frac{\hbar^2}{2m_b^*}k_{\sigma}^2+ \vartheta_{0}\right) \mathbb{\hat{I}}+\beta\left(\hat{\tau}_y\,k_{y\sigma}-\hat{\tau}_x\,k_{x\sigma}\right)k_{z\sigma}^2\nonumber\\
  && +\,\alpha\left(\hat{\tau}_x\,k_{y\sigma}-\hat{\tau}_y\,k_{x\sigma}\right),
  \label{2}
\end{eqnarray} where $\vartheta_{0} = E_F + V_0$. The subscript $\sigma$ indicates whether the spin $\mathbf{s_\sigma^\pm}$ is parallel ($\sigma=\uparrow$) or antiparallel ($\sigma=\downarrow$) to $\mathbf{n}_l$; however, in algebraic expressions it takes the values $+1$ and $-1$, respectively. In addition, the superscript $\pm$ represents the spin-up $(+)$ and spin-down $(-)$ states arising from SOC. Here, $m_b^*$ is the effective mass in the SC and $k_\sigma$ is the electron wave vector in the barrier. (see Suppl. Mater., Sec. A, Fig. S1). Finally, $\beta$ and $\alpha$ are the \emph{Dresselhaus} and \emph{Rashba} SOC constants, respectively.

\begin{figure}
\includegraphics[width=0.45\textwidth]{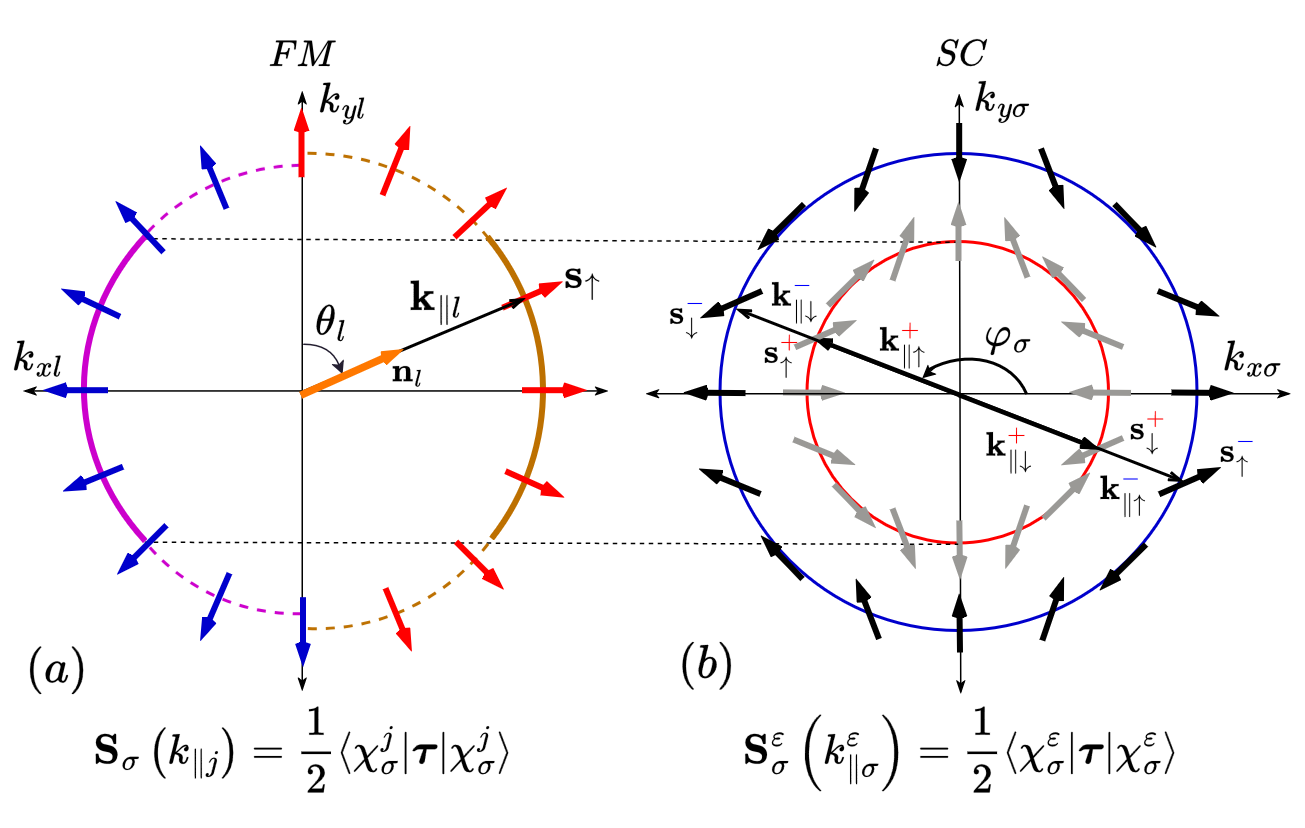}
\caption{\label{F2} Spin orientation corresponding to the spinors $|\chi_\sigma^j\rangle$ and $|\chi_\sigma^\varepsilon\rangle$ in the $k_{xl} \text{-} k_{yl}$ and $k_{x\sigma} \text{-} k_{y\sigma}$ planes for FM and SC, respectively, in the case where both are parallel to the barrier plane. The orientation of the wave vector $k_{\|l}$ is related to $k_{\|\sigma}^{+}$ (spin-up state) and $k_{\|\sigma}^{-}$ (spin-down state) by $\varphi_{\sigma}^{\varepsilon} = \theta_l + \varepsilon\sigma\frac{\pi}{2}$. Moreover, $\mathbf{s}_{\sigma}$ and $\mathbf{s}_{\sigma}^{\varepsilon}$ represent spins that are elements of $\mathbf{S}_{\sigma}$ and $\mathbf{S}_{\sigma}^{\varepsilon}$, respectively.}
\end{figure}

The Hamiltonian $\mathcal{\hat{H}}_{\sigma}^\varepsilon$ has energy eigenvalues,
\begin{equation}
  E_\sigma^\varepsilon = \frac{\hbar^2}{2m_b^*}k_{\sigma}^2+\vartheta_{0}+\varepsilon\, k^\varepsilon_{\|\sigma}J_{k_{z\sigma}},
  \label{5}
\end{equation} where $J_{k_{z\sigma}} = \beta\,k_{z\sigma}^2 - \alpha\,\sin\left(2\varphi_\sigma\right)$, with $\varepsilon=\pm$. The spinor that diagonalizes the Hamiltonian is given by \cite{Perel03}:
\begin{equation}
 |\chi_\sigma^\varepsilon\rangle=\frac{1}{\sqrt{2}}\left(
     \begin{matrix}
     1 \\
     -\varepsilon e^{-i\varphi_\sigma}\\
     \end{matrix}
     \right).
     \label{6}
\end{equation} where $\varphi_\sigma$ is related to the direction $\mathbf{n}_l$ (see Fig.~\ref{F2}) through $\varphi_\sigma^\varepsilon = \theta_l + \varepsilon\sigma\frac{\pi}{2}$ (see Suppl. Mater., Sec. B).

\subsection{Solution of {\em Schrödinger-Pauli} equation}

The {\em Schrödinger-Pauli} equation, $\hat{\mathcal{H}}|\psi\rangle = E|\psi\rangle$, where $\hat{\mathcal{H}}$ is defined in terms of the step function $\Theta\left(z\right)$ as:
\begin{equation}
 \hat{\mathcal{H}} = \mathcal{\hat{H}}_l\Theta\left(-z\right)+\mathcal{\hat{H}}_{\sigma}^\varepsilon\Theta\left(z\right)\Theta\left(a-z\right)+\mathcal{\hat{H}}_r\Theta\left(z-a\right),
 \label{7}
\end{equation}
whose solution can be decoupled into a plane wave in the direction of $\bm{\kappa} _{\|}$, which can take the values $\mathbf k_{\|j}$ and $\mathbf k^\varepsilon_{\|\sigma}$, corresponding to the FM and SC regions, respectively. This solution is modulated by a wave function $|\varPhi_\sigma^\varsigma(z)\rangle = u_\sigma^\varsigma\left(z\right)|\chi_\sigma^\varsigma\rangle$ (with $\varsigma = j,\varepsilon$), which is the product of a periodic function and its respective spinor. That is,
$|\psi\rangle = e^{i\,\bm{\kappa} _{\|}\cdot\boldsymbol{r}\left(x,y\right)}u_\sigma^\varsigma\left(z\right)|\chi_\sigma^\varsigma\rangle$,
where $\boldsymbol{r}\left(x,y\right)$ is a position vector on the barrier plane, parallel to the $(0\,0\,1)$ plane. Then, $|\varPhi_\sigma^\varsigma(z)\rangle$ is determined by the following expression:
\begin{equation}
\left\{
       \begin{array}{lcc}
             \frac{1}{2\sqrt{k_\sigma^l}}\,e^{ik_\sigma^lz}|\chi_{\sigma}^l\rangle+\mathcal{R}_{\sigma}e^{-ik_\sigma^lz}|\chi_{\sigma}^l\rangle & if & z < 0\\
            \\\displaystyle\sum_{\varepsilon=\pm}\left[\left(\mathcal{C}_\sigma^\varepsilon\,e^{i\varrho_\varepsilon z}+\mathcal{D}_\sigma^\varepsilon\,e^{-i\varrho_\varepsilon z}\right)|\chi_{\sigma}^\varepsilon\rangle\right] &  if & 0 < z < a, \\
            \\\mathcal{T}_{\sigma}e^{ik_\sigma^rz}|\chi_{\sigma}^r\rangle  &  if  & z > a
        \end{array}
                \right.
 \label{8}
\end{equation} where $k_\sigma^j = k_\sigma^j(\theta,k_{\|\,j})$ and $\varrho_\varepsilon = i\rho_\varepsilon(\theta_l,k_{\|\sigma}^\varepsilon)$ represent the magnitude of the wave vectors for the spins in the layers $\mathrm{L}_j$ and within the barrier, respectively (see Suppl. Mater., Sec. C). Consequently,

\begin{equation}
  k_\sigma^j = \sqrt{\frac{2m_0}{\hbar^2}\left(E+\sigma\frac{\Delta^j_{xs}}{2}\cos\left(\theta \right)\right)-k_{\|\,j}^2},
  \label{9}
\end{equation} where $m_0$ is the free-electron mass and $E$ denotes the energy of the electron. The wave vector $k_\sigma^l$ is obtained when $\theta = 0$ in the previous expression, whereas the wave vector $k_\sigma^r$ takes two values: one that coincides with $k_\sigma^l$ (for $\theta = 0$) and another corresponding to $\theta = \pi$. From now on, we set $k_{\|\,l} = k_{\|\,r} \equiv k_{\|}$.

\begin{equation}
 \rho_\varepsilon = \sqrt{\frac{\frac{2m_b^*}{\hbar^2}\left(V_{eff}-E\right)+\left(k_{\|\sigma}^\varepsilon\right)^2-\varepsilon\,\frac{2m_b^{*}}{\hbar^2}\alpha\,k_{\|\sigma}^\varepsilon \sin\left(2\theta_l\right)}{\mu_\sigma^\varepsilon}},
\label{10}
 \end{equation} where $\mu_\sigma^\varepsilon = 1+\varepsilon\frac{2m_b^{*}}{\hbar^2}\beta\,k_{\|\sigma}^\varepsilon$.

\subsection{Transmission probability}

The coefficients $\mathcal{R}_\sigma$, $\mathcal{C}_\sigma^\varepsilon$, $\mathcal{D}_\sigma^\varepsilon$, and $\mathcal{T}_{\sigma}$ in equation (\ref{8}) are determined applying boundary conditions at $z_l = 0$ and $z_r = a$. That is \cite{Glazov05, Li11},
\begin{equation}
 \begin{aligned}
  |\varPhi_{\sigma}^j(z_j)\rangle =&\,|\varPhi_{\sigma}^\varepsilon(z_j)\rangle\\
  \frac{1}{m_j^*}\partial_z|\varPhi_{\sigma}^j(z)\rangle \Big|_{z = z_j} =&\,\frac{1}{m_\varepsilon^*}\partial_z|\varPhi_{\sigma}^\varepsilon(z)\rangle \Big|_{z = z_j}
 \end{aligned}
\end{equation} where $m_\varepsilon^* = \left(\mu_\sigma^\varepsilon\right)^{-1}m_b^*$.

Performing algebraic processes and considering the notation $q_\sigma^\varepsilon = \mu_\sigma^\varepsilon\rho_\varepsilon$ (see Suppl. Mater., Sec. D) we obtain the following expression:
\begin{equation}
 \Lambda_\pm = \dfrac{\kappa_l}{\sqrt{k_\sigma^l}}\dfrac{\left( q_\sigma^\varepsilon \pm i\kappa_r  \right)}{\left(\kappa_l+\kappa_r\right)q_\sigma^\varepsilon\,\lambda_\sigma^\varepsilon}e^{\pm \rho_\varepsilon a},
 \label{11}
\end{equation}
where $\lambda_\sigma^\varepsilon = \cosh\left(\rho_\varepsilon\,a\right)-i\dfrac{\kappa_l\kappa_r-\left(q_\sigma^\varepsilon\right)^2}{\left(\kappa_l+\kappa_r\right)q_\sigma^\varepsilon}\sinh\left(\rho_\varepsilon\,a\right)$ and $\kappa_j= \chi_b\,k_{\sigma}^j(\theta,k_{\|})$, with $\chi_b$ the relative effective mass of the SC. It then follows that $\mathcal{C}_\sigma^\varepsilon = \Lambda_+$ and $\mathcal{D}_\sigma^\varepsilon = \Lambda_-$. which implies that the transmission coefficient is given by:
\begin{equation}
   \mathcal{T}_{\sigma}^\varepsilon =\frac{1}{\sqrt{k_\sigma^l}}\frac{\kappa_l}{\kappa_l+\kappa_r}e^{\text{-}i\,k_\sigma^r\,a}\displaystyle\sum_{\varepsilon=\pm}\left(\lambda_\sigma^\varepsilon\right)^{\text{-}1}.
\label{13}
\end{equation}
Consequently, the expression obtained for the transmission probability is:
\begin{equation}
   \mathrm{T}_\sigma\left(k_{\|},\theta\right)=\frac{\kappa_l\,\kappa_r}{\left(\kappa_l+\kappa_r\right)^2}\left|\displaystyle\sum_{\varepsilon=\pm}\left(\lambda_\sigma^\varepsilon\right)^{\text{-}1}\right|^2.
   \label{14}
\end{equation}

\subsection{Tunnel Magnetoresistance}

The TMR = $(G_P - G_{AP})/G_{AP}$, where the conductance $G_P$ occurs when $\mathbf{n}_l \parallel \mathbf{n}_r$ and the conductance $G_{AP}$ occurs when $\mathbf{n}_l \nparallel\mathbf{n}_r$. According to the \emph{Landauer-B\"{u}ttiker} formula for T $\approx$ 0 K \cite{Wim09}:
\begin{equation}
 G_\ell =\frac{e^2\,A_c}{\left(2\pi\right)^3\hbar}\int_{k_{\|_0}}^{k_{\|_{max}}}dk_{\|}\,\mathbb{T}\left(k_{\|},\theta\right),
 \label{15}
\end{equation} where $\ell = P$ if $\theta = 0$ and for $\ell = AP$, $\theta = \pi$; $A_c$ is the cross-section area of the junction, and $\mathbb{T}(k_{\|},\theta) = \mathrm{T}_\uparrow\left(k_{\|},\theta\right)+\mathrm{T}_\downarrow\left(k_{\|},\theta\right)$  is the transmittance coefficient.

\section{Results and Discussions}

\begin{figure}[H]
\includegraphics[width=0.47\textwidth]{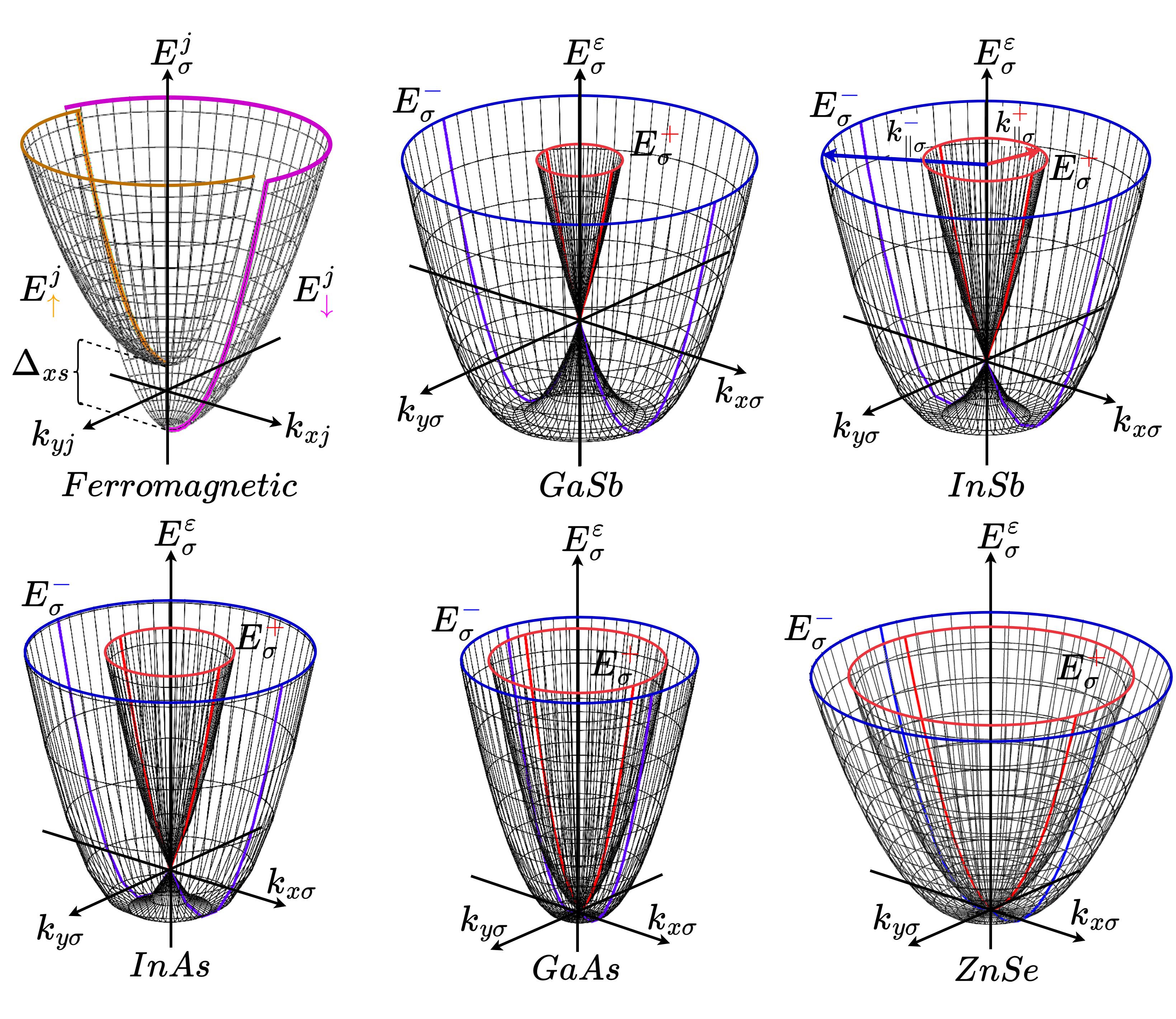}
\caption{\label{F3} Schematic representation of the k-space in 2D for the energy eigenvalues $E_{\sigma}^j$ corresponding to FM materials and $E_{\sigma}^{\varepsilon}$ related to SCs with \emph{Dresselhaus} and \emph{Rashba} SOC. For this analysis $E = E_F$, $\theta_m = \pi/4$, $\mathbf{n}_l \parallel [1\,1\,0]$ and $k_{z\sigma} = 0.44\,\text{\AA}^{-1}$ was considered.}
\end{figure}

For the analysis of the TMR in the PSV under study, we assume that $k_{\|\sigma}^+ = k_{\|}$, and that the condition $E_\sigma^+ = E_\sigma^-$ (see Fig. \ref{F3}) determines $k_{\|\sigma}^-$, given by:
\begin{equation}
 k_{\|\sigma}^- = \frac{2m_b^*}{\hbar^2}J_{k_{z\sigma}} + k_{\|},
 \label{16}
\end{equation} as detailed in the (see Suppl. Mater., Sec. E) Consequently, the wave vectors parallel to the barrier plane will be constrained by the condition $k_{\|}^2 \leq min\left\{\frac{2m_0}{\hbar^2}\left(E-\frac{\Delta^j_{xs}}{2}\right)\right\}$ (see expression (\ref{9})). For the case $E = E_F$, the lowest value of $k_{\|}$ is given by $\mathrm{Fe}_{90}\mathrm{Cr}_{10}$, according to the Table \ref{Tab1}. In our calculations, we consider $k_{\|_0} = 0.1\, \mathrm{nm}^{-1}$ and $k_{\|_{max}} = 1\,\mathrm{nm}^{-1}$, based on various theoretical reports \cite{Perel03, Glazov05, Li11, Lu12}. On the other hand, the condition $\mu_\sigma^- > 0$ constrains expression (\ref{15}), thereby allowing the parameterization of $k_{z\sigma}$ for $k_{\|} = k_{\|_{max}}$. That is, the maximum value of the parameter $k_{z\sigma}^{max}$ is given by:
\begin{equation}
 k_{z\sigma}^{max} < \left(\frac{\hbar^2}{2m_b^*\beta}\right)^2-\frac{\hbar^2}{2m_b^*\beta}k_{\|} + \frac{\alpha}{\beta}\sin\left(2\theta_l\right),
\end{equation} values shown in Table \ref{Tab2}.

\begin{table}[H]
\caption{\label{Tab1} Fermi wave vectors of majority ($k_{F\uparrow}$) and minority ($k_{F\downarrow}$) spins are given in units of $\text{\AA}^{-1}$, while $\Delta_{xs}^j$ is reported in eV, like $E_F$. For $\mathrm{Fe}_{90}\mathrm{Cr}_{10}$ (bcc) and Co (fcc), the values of $\Delta_{xs}^j$ were obtained by interpolation using references \cite{Korz09} and \cite{liza17}, respectively. For the remaining cases, the values were calculated following the procedure of N. N. Beletskii \emph{et al}.\cite{Bel07}. Data for Fe and $\mathrm{Ni}_{80}\mathrm{Fe}_{20}$ correspond to the $(0\,1\,1)$ direction averaged over the three crystal directions \cite{Grun01}.}
\begin{ruledtabular}
\begin{tabular}{ccccc}
                                    & $k_{F\uparrow}$  & $k_{F\downarrow}$  & $E_F$             & $\Delta_{xs}^j$ \\ \hline
$\mathrm{Fe}_{90}\mathrm{Cr}_{10}$  & ${}^{\dag}$1.10  & ${}^{\dag}$0.32    & ${}^{\ddag}$2.50  & 4.22 \\
Fe                                  & 1.05             & 0.44               & 2.48              & 3.46 \\
Co                                  & ${}^{\dag}$0.94  & ${}^{\dag}$0.62    & ${}^{\ddag}$2.40  & 1.90 \\
$\mathrm{Ni}_{80}\mathrm{Fe}_{20}$  & 1.05             & 0.88               & 3.59              & 1.24 \\
Ni                                  & 1.00             & 0.97               & 3.99              & 0.94
\end{tabular}
\end{ruledtabular}
\begin{flushleft}
\footnotesize
${}^{\dag}$Calculated value and ${}^{\ddag}$Estimated value.
\end{flushleft}
\end{table}

Considering the previous conditions, for the results in which the efficiency of the PSV structures under study is calculated as a function of the SC thickness and/or the effect of the \emph{Dresselhaus} and \emph{Rashba} SOC, we use $E = E_F$, $\theta_m = \pi/4$, $\mathbf{n}_l \| [1\,1\,0]$ and $k_{z\sigma} = 0.44,\text{\AA}^{-1}$ is used. This value corresponds to the smallest one among the SCs considered (see Table \ref{Tab2}), allowing a direct comparison among all possible PSV combinations analyzed (a total of 125). Unless otherwise specified, these values are used throughout this work.

Fig. \ref{F3} shows the 2D $k$-space distribution of the energy eigenvalues, illustrating the effect described by Eq. (\ref{16}) for the SC materials, which appears as radial patterns. Moreover, the minimum energy values for each SC decrease in magnitude as $k_{\|}^{+}$ increases, for $\varphi_{\uparrow}^{-} = -\pi/4$, as follows: 3.58 eV for GaSb, 1.62 eV for InSb, 1.03 eV for InAs, 0.13 eV for GaAs, and 0.08 eV for ZnSe. These variations directly influence the effective barrier potential. On the other hand, the SOC constants for zinc-blende (III–V) semiconductors such as GaSb, InSb, InAs, and GaAs are listed in Table \ref{Tab2}, similarly to those of the ZnSe compound, which belongs to the (II–VI) group. In addition, the barrier heights are defined as follows: 0.75 eV for GaSb and GaAs \cite{Zenger04, Autes10, Kondo12}, and 0.32 eV for InAs \cite{Lu12}. For InSb and ZnSe, values of 0.20 eV and 1.90 eV are estimated in this work, respectively.

\begin{table}[H]
\caption{\label{Tab2} The relative effective mass \cite{Fabian07}, band gap (in eV) \cite{Wink03}, and the SOC constants were used in the calculations. The \emph{Dresselhaus} ($\beta$) and \emph{Rashba} ($\alpha$) parameters are expressed in units of [eV$\text{\AA}^3$] and [eV$\text{\AA}$], respectively \cite{Perel03, Lu10, Dakh20}. Additionally, the approximate maximum values of $k_{z\sigma}$ are indicated, in units of $(\text{\AA}^{-1})$, for the case when $k_{\|} = k_{\|_{max}}$. }
\begin{ruledtabular}
\begin{tabular}{cccccc}
SCs   & $\chi_b$       & $E_g$         & $\beta$      & $\alpha$       & $k_{z\sigma}$\\ \hline
GaSb  & 0.041          & ${}^{b}$0.810 & 187          & 0.30           & 0.443 \\
InSb  & 0.0135         & 0.237         & 220          & 0.12           & 1.232 \\
InAs  & 0.023          & 0.418         & 130          & 1.01           & 1.221 \\
GaAs  & 0.067          & 1.519         & ${}^{a}$27.6 & ${}^{c}$0.0873 & 2.010 \\
ZnSe  & ${}^{a}$0.160  & 2.820         & ${}^{a}$14.3 & 0              & 1.615
\end{tabular}
\end{ruledtabular}
\begin{flushleft}
\footnotesize
${}^{a}$Ref. \cite{Wink03}, ${}^{b}$Ref. \cite{Fabian07} and ${}^{c}$Ref. \cite{John08}.
\end{flushleft}
\end{table}

\begin{figure}[H]
\centering
\includegraphics[width=0.49\textwidth]{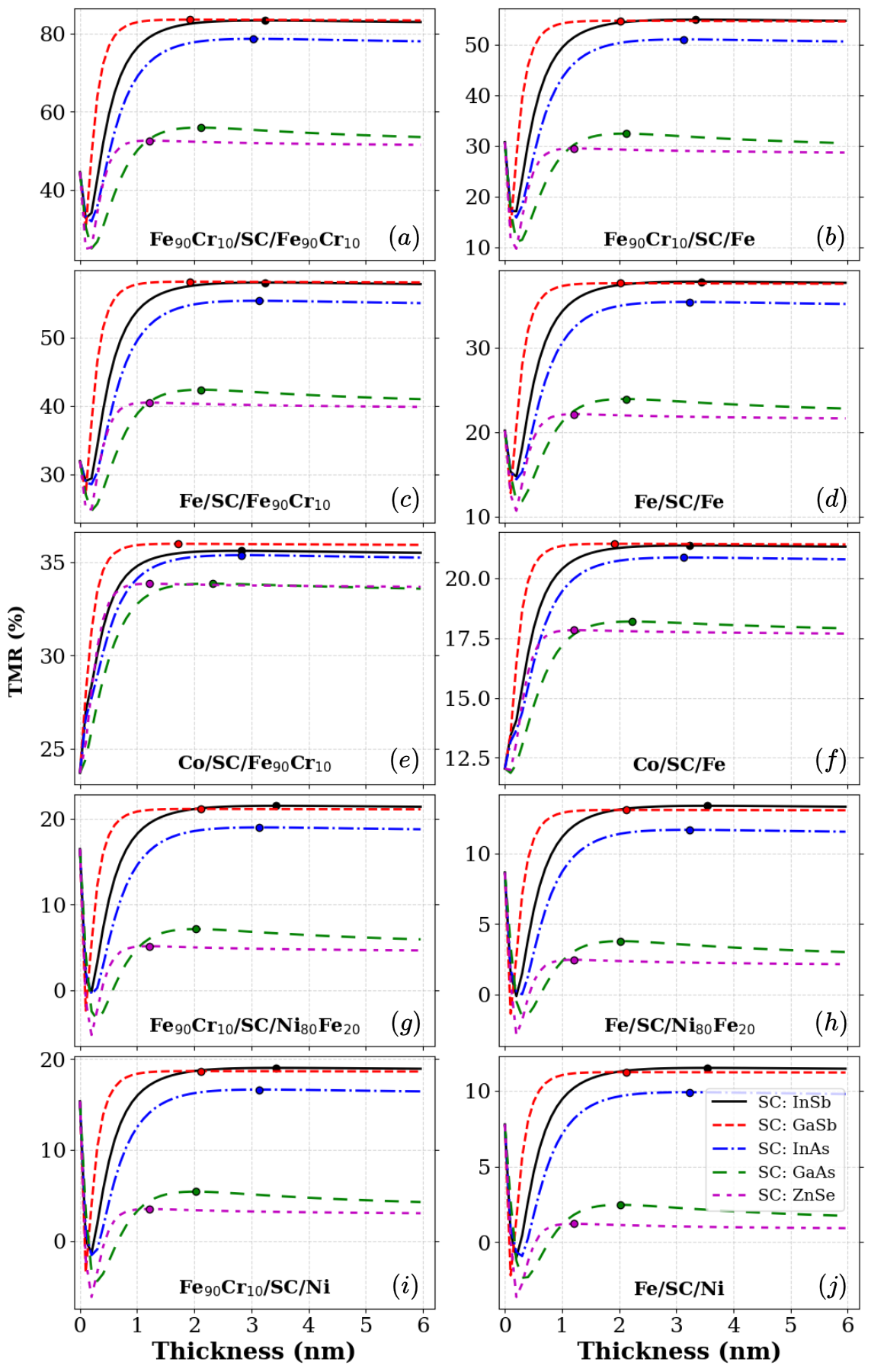}
\caption{\label{F4} Calculated TMR for the L\textsubscript{l}/SC/L\textsubscript{r} PSV structures as a function of the SC layer thickness for $E = E_F$, $\theta_m = \pi/4$, $\mathbf{n}_l \| [1\,1\,0]$ and $k_{z\sigma} = 0.44\,\text{\AA}^{-1}$. Results are shown for InSb (black solid line), GaSb (red dashed line), InAs (blue dash-dotted line), GaAs (green custom dashed pattern), and ZnSe (magenta custom compound dashed pattern).}
\end{figure}

In Fig.\ref{F4}, the PSV structures exhibiting the highest TMR values are shown, while the remaining ones are provided in the Suppl. Mater. (Sec. F, Fig. S9). In (a), the $\mathrm{Fe}_{90}\mathrm{Cr}_{10}$/SC/$\mathrm{Fe}_{90}\mathrm{Cr}_{10}$ PSV with the highest TMR of 83.60\% is observed for the GaSb spacer with $a = 1.92$ nm, followed by the InSb spacer with $a = 3.23$ nm and a TMR of 83.38\%. Table \ref{Tab3}, associated with panels (a)–(e) of the figure, lists the maximum TMR values and the corresponding spacer thicknesses at which they occur, including an ascending order (Rk.) according to the TMR value.

Continuing the analysis of Fig. \ref{F4}(b) and (c), together with Table \ref{Tab3} for the corresponding PSV structures and Fig. S9, it is observed that the TMR value of the PSV L\textsubscript{l}/SC/L\textsubscript{r} differs from that of L\textsubscript{r}/SC/L\textsubscript{l}. In other words, the FM layers do not switch in the PSV structure

\begin{figure}[H]
\includegraphics[width=0.48\textwidth]{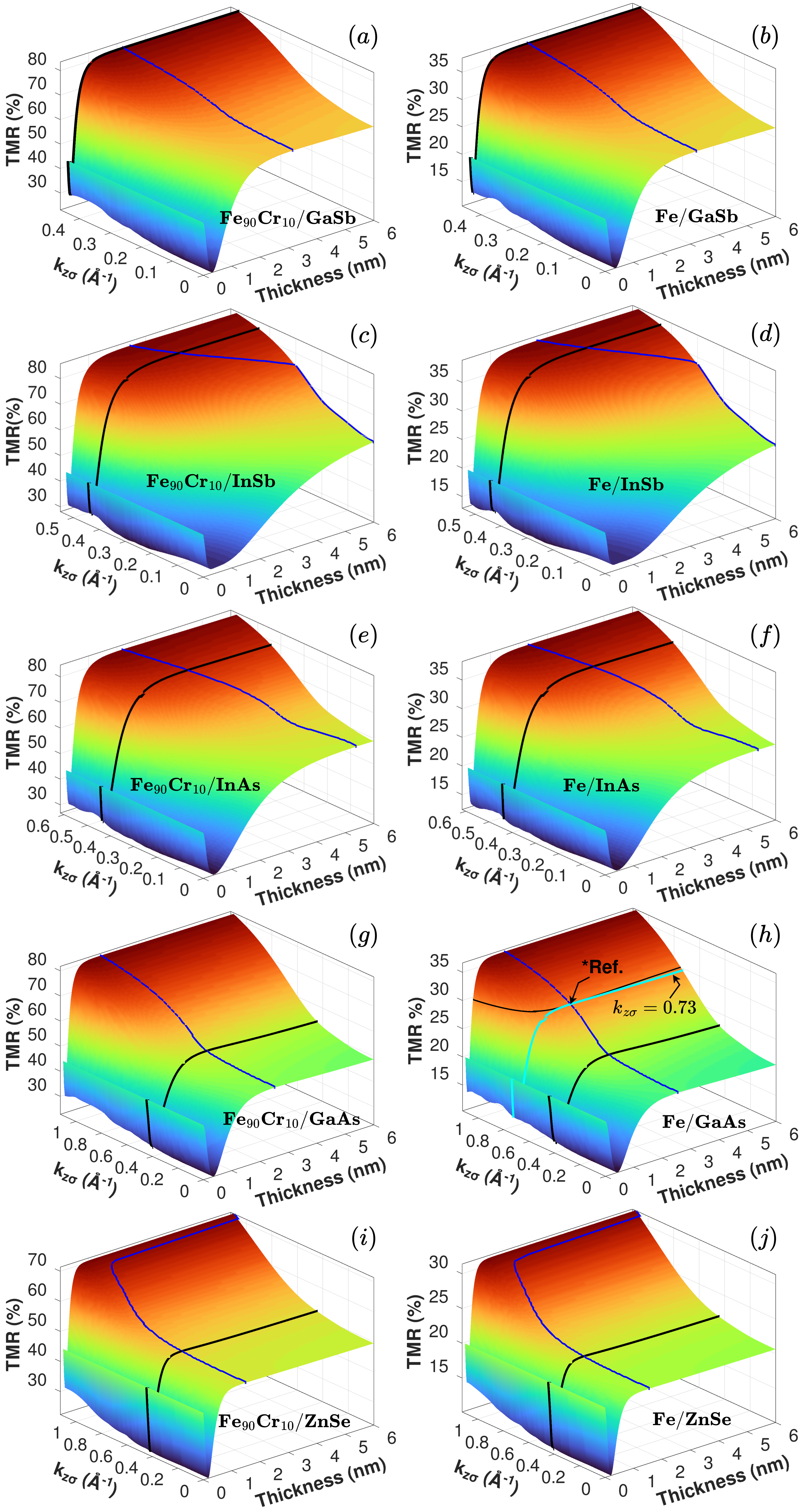}
\caption{\label{F5} TMR for the PSV $\mathrm{L}_k$/SC (equivalent to $\mathrm{L}_k$/SC/$\mathrm{L}_k$), as a function of the SC thickness and the wave vector $k_{z\sigma}$ for $E = E_F$, $\theta_m = \pi/4$ and $\mathbf{n}_l \| [1\,1\,0]$. The thick black curve correspond to the $k_{z\sigma}$ value defined in this work , the blue curve indicates the ridge of the surface. *Ref. \cite{Autes10} (TMR = 30\%).}
\end{figure}

Fig.\ref{F5} shows the TMR for the PSV L\textsubscript{k}/SC (equivalent to L\textsubscript{k}/SC/L\textsubscript{k}), with $\mathrm{L}_{\mathrm{k}} = \{\mathrm{Fe}_{90}\mathrm{Cr}_{10}, \mathrm{Fe}\}$, as a function of the SC thickness and the wave vector $k_{z\sigma}$. The thick black curve correspond to the $k_{z\sigma}$ value defined in this work and represent the results shown in Fig.\ref{F4}(d). The blue curve indicates the ridge of the surface, i.e., the maximum TMR values achieved. It is worth noting that the intersection between these two curves corresponds to the maximum TMR value listed in Table \ref{Tab3}.

Consequently, for panels (a) and (b), corresponding to the GaSb spacer, the TMR as a function of thickness lies along the edge of the surface. For the remaining cases, $k_{z\sigma}$ varies up to values that do not exceed those listed in Table \ref{Tab2}. In this way, in panels (c)–(f), it is observed that the TMR value increases as the spacer thickness decreases, a behavior that is not observed in panels (i) and (j), where, for small thickness variations (between 1 and 2 nm), the TMR initially increases and then, for $k_{z\sigma} > 1.1\,\text{\AA}^{-1}$, tends to saturate. In general, it is observed that as $k_{z\sigma}$ increases, the TMR also increases, since this parameter is directly related to $E_D$ (see Eq. (\ref{5})), and consequently to an increase in $V_{eff}$.

On the other hand, in panel (h), for the PSV Fe/GaAs/Fe with $k_{z\sigma} = 0.73\,\text{\AA}^{-1}$, a TMR of 30\% is obtained, a value reported by G. Autès \emph{et al}. \cite{Autes10}. However, in that study, this result was obtained under different conditions ($\Delta_{xs} = 0$).

\begin{table*}[t]
\caption{\label{Tab3} Calculation of the maximum TMR (in \%) and the corresponding SC thickness $a$ (in nm) forming the PSV. Rk denotes the descending order according to the TMR value. For this analysis $\theta_m = \pi/4$, $\mathbf{n}_l \| [1\,1\,0]$ and $k_{z\sigma} = 0.44\,\text{\AA}^{-1}$ were considered.}
\begin{ruledtabular}
\begin{tabular}{c|ccccccccccccccc}
    & \multicolumn{3}{c}{\textbf{GaSb}}                                    & \multicolumn{3}{c}{\textbf{InSb}}                                    & \multicolumn{3}{c}{\textbf{InAS}}                                    & \multicolumn{3}{c}{\textbf{GaAs}}                                    & \multicolumn{3}{c}{\textbf{ZnSe}}               \\ \cline{2-16}
PSV & Rk.                & \textbf{TMR} & \multicolumn{1}{c|}{$\boldsymbol{a}$} & Rk.                & \textbf{TMR} & \multicolumn{1}{c|}{$\boldsymbol{a}$} & Rk.                & \textbf{TMR} & \multicolumn{1}{c|}{$\boldsymbol{a}$} & Rk.                & \textbf{TMR} & \multicolumn{1}{c|}{$\boldsymbol{a}$} & Rk.                & \textbf{TMR} & $\boldsymbol{a}$ \\ \hline
\multicolumn{1}{c|}{$\mathrm{Fe}_{90}\mathrm{Cr}_{10}$/SC/$\mathrm{Fe}_{90}\mathrm{Cr}_{10}$} & {\footnotesize 1}  & 83.60        & \multicolumn{1}{c|}{1.92}       & {\footnotesize 2}  & 83.38        & \multicolumn{1}{c|}{3.23}       & {\footnotesize 3}  & 78.68        & \multicolumn{1}{c|}{3.03}       & {\footnotesize 6}  & 55.98        & \multicolumn{1}{c|}{2.12}       & {\footnotesize 10} & 52.64        & 1.21       \\
\multicolumn{1}{c|}{Fe/SC/$\mathrm{Fe}_{90}\mathrm{Cr}_{10}$} & {\footnotesize 4}  & 58.21        & \multicolumn{1}{c|}{1.92}       & {\footnotesize 5}  & 58.09        & \multicolumn{1}{c|}{3.23}       & {\footnotesize 7}  & 55.43        & \multicolumn{1}{c|}{3.13}       & {\footnotesize 12} & 42.40        & \multicolumn{1}{c|}{2.12}       & {\footnotesize 13} & 40.52        & 1.21       \\
\multicolumn{1}{c|}{$\mathrm{Fe}_{90}\mathrm{Cr}_{10}$/SC/Fe} & {\footnotesize 9}  & 54.74        & \multicolumn{1}{c|}{2.02}       & {\footnotesize 8}  & 54.96        & \multicolumn{1}{c|}{3.33}       & {\footnotesize 11} & 51.06        & \multicolumn{1}{c|}{3.13}       & {\footnotesize 22} & 32.46        & \multicolumn{1}{c|}{2.12}       & {\footnotesize 23} & 29.54        & 1.21       \\
\multicolumn{1}{c|}{Fe/SC/Fe} & {\footnotesize 15} & 37.60       & \multicolumn{1}{c|}{2.02}       & {\footnotesize 14} & 37.79        & \multicolumn{1}{c|}{3.43}       & {\footnotesize 18} & 35.39        & \multicolumn{1}{c|}{3.23}       & {\footnotesize 27} & 23.90        & \multicolumn{1}{c|}{2.12}       & {\footnotesize 28} & 22.11        & 1.21       \\
\multicolumn{1}{c|}{Co/SC/$\mathrm{Fe}_{90}\mathrm{Cr}_{10}$} & {\footnotesize 16} & 35.98        & \multicolumn{1}{c|}{1.72}       & {\footnotesize 17} & 35.61        & \multicolumn{1}{c|}{2.83}       & {\footnotesize 19} & 35.37        & \multicolumn{1}{c|}{2.83}       & {\footnotesize 20} & 33.85        & \multicolumn{1}{c|}{2.32}       & {\footnotesize 21} & 33.84        & 1.21
\end{tabular}
\end{ruledtabular}
\end{table*}

\begin{figure}[h]
\includegraphics[width=0.48\textwidth]{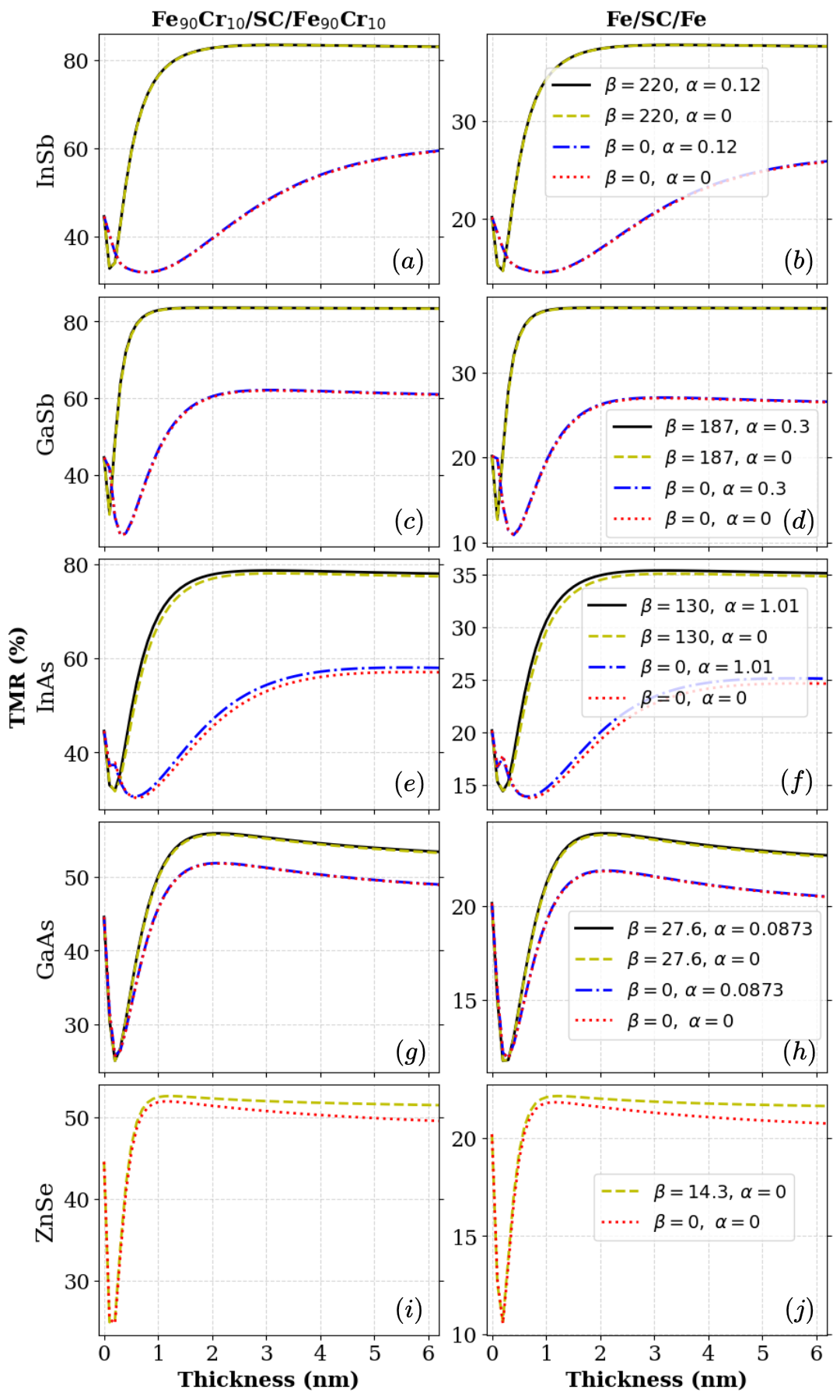}
\caption{\label{F6} Calculated TMR for the $\mathrm{L}_k$/SC/$\mathrm{L}_k$ PSV structures as a function of the SC layer thickness and the \emph{Dresselhaus} ($\beta$) [eV$\text{\AA}^3$] and \emph{Rashba} ($\alpha$) [eV$\text{\AA}$] SOC parameters, for $E = E_F$, $\theta_m = \pi/4$, $\mathbf{n}_l \| [1\,1\,0]$ and $k_{z\sigma} = 0.44\,\text{\AA}^{-1}$. Results are shown with both $\beta$ and $\alpha$ (black solid line), with $\beta$ and without $\alpha$ (green dashed line), without $\beta$ and with $\alpha$ (blue dash-dotted line), and without both $\beta$ and $\alpha$ (red dotted line).}
\end{figure}

As in Fig. \ref{F5}, Fig. \ref{F6} analyzes the same PSV structures; however, in this case, the effect of SOC on the TMR is shown. In panels (a)–(j), it is clearly observed that the Rashba SOC has no significant impact due to its small magnitude (see Table \ref{Tab2}). In contrast, the Dresselhaus SOC produces a notable enhancement of the TMR exceeding 40\% in the PSV structures shown in panels (a) and (c), and over 30\% in (e). Likewise, panels (b) and (d) exhibit an increase greater than 20\%, while in (f) the enhancement reaches approximately 15\%. These results can be corroborated in Suppl. Mater. (Sec. F, Fig. S10). Furthermore, for PSVs with a GaAs spacer, the TMR increase does not exceed 4\%, and it is below 2\% in the case of ZnSe. It should be noted that, according to the value of $k_{z\sigma}$ considered in this study, only 19.36\% of the total \emph{Dresselhaus} SOC constant is taken into account, as given by Eq. \ref{5}.

\begin{figure}[H]
\includegraphics[width=0.48\textwidth]{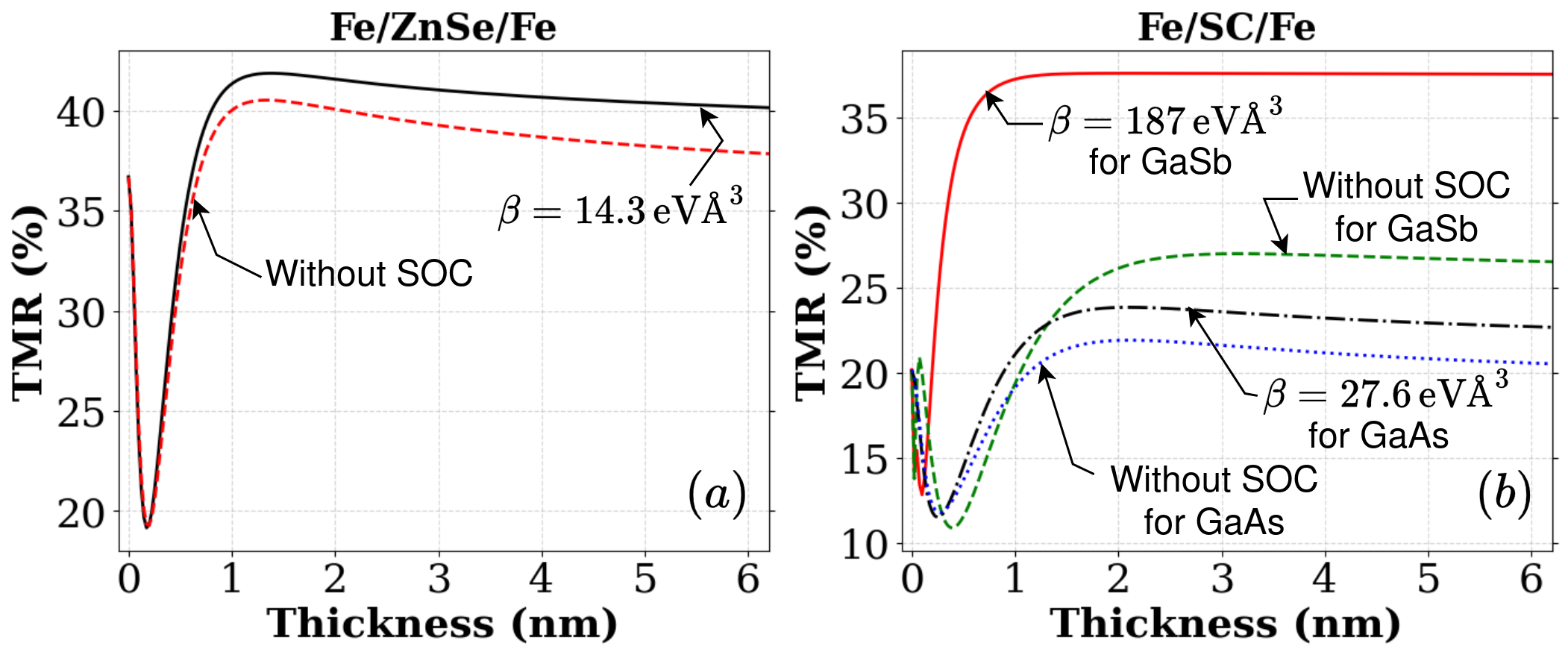}
\caption{\label{F7} Calculation of the TMR as a function of the SC thickness for the PSV structures: (a) Fe/ZnSe/Fe, with (black solid line) and without (red dashed line) \emph{Dresselhaus} SOC; (b) Fe/SC/Fe with SC = GaSb (red solid line: with \emph{Dresselhaus} SOC; green dashed line: without \emph{Dresselhaus} SOC) and SC = GaAs (black dash-dot line: with \emph{ Dresselhaus} SOC; blue dotted line: without \emph{Dresselhaus} SOC). In both panels the parameters $E = E_F$, $\theta_m = \pi/4$, $\mathbf{n}_l \| [0\,1\,0]$ and $k_{z\sigma} = 0.44\,\mathring{A}^{-1}$ were used.}
\end{figure}

Next, Fig. \ref{F7} shows that for $\theta_m = \pi/4$ and $\mathbf{n}_l \| [0\,1\,0]$, in panel (a) the PSV Fe/ZnSe/Fe structure reaches a maximum TMR of 41.88\% (for $a = 1.35$ nm) with \emph{Dresselhaus} SOC, and 40.53\% (for $a = 1.33$ nm) without \emph{Dresselhaus} SOC. The latter value is in reasonable agreement with the results obtained from \emph{ab initio} calculations and Green’s functions reported by J. Peralta-Ramos \emph{et al}. \cite{Peralta06}, who suggest $\Delta_{xs} = 2.552$ eV for Fe. In our model, the remaining fitted parameters mentioned in that work were $V_{0} = 1.1$ eV and a \emph{ Fermi} level of $E_F = 1.57$ eV, experimental values reported by M. Eddrief \emph{et al}. \cite{Edd02}.

In the same figure, panel (b) compares our results with those obtained by K. Kondo \cite{Kondo12}, under the same parameters defined in this work except for the value of $\theta_m$. The following discrepancies are found: for the Fe/GaAs/Fe PSV, a maximum TMR of 23.9\% (see Table \ref{Tab3}) is obtained with \emph{Dresselhaus} SOC, which differs from Kondo’s report where the TMR converges to $-60\%$ from $a = 0.75$ nm. In the absence of \emph{Dresselhaus} SOC, the TMR reaches approximately 21\%, a value very close to that obtained in this work (21.9\%) for a thickness of 2.2 nm. Finally, for the Fe/GaSb/Fe PSV, Kondo’s study reports a TMR that converges to 140\% from 2.5 nm with \emph{Dresselhaus} SOC, a result that does not agree with the value calculated in this study, as shown in Table \ref{Tab3}, where a considerable difference is observed. So far, no other references have reported a value similar to that given by K. Kondo for the mentioned PSV.

\section{Conclusions}

The theoretical analysis shows that the \emph{Dresselhaus} SOC present in the SCs under study, which is related to the magnitude of the wave vector $k_{z\sigma}$ (in agreement with the work of S. D. Ganichev \emph{et al}. \cite{Gani04}), determines the dependence of the TMR on the SC thickness in the various PSV structures analyzed. This occurs because both the intrinsic effect of the SC material and the associated external factor contribute to an increase in the TMR. Moreover, the presence of \emph{Rashba SOC} introduces an additional contribution that depends on the direction of $\mathbf{n}_l$ and is independent of $\theta_m$. Consequently, these two specific cases confirm that the SOC is subject to gate control \cite{Aws02}, placing it among the key mechanisms that bridge the physics of spin–orbit coupling with modern spintronics \cite{Igor04, Manc15}.

The computational results of the model proposed in this work indicate that, when the condition $E_F^{l} \leq E_F^{r}$ is fulfilled, regardless of whether $\Delta_{xs}^{l} < \Delta_{xs}^{r}$ holds or not, the maximum TMR of the PSV structures L\textsubscript{l}/SC/L\textsubscript{r} is larger than that of L\textsubscript{r}/SC/L\textsubscript{l} for $\mathrm{SC}=\{\mathrm{GaSb},\,\mathrm{InSb},\,\mathrm{InAs}\}$, presumably because these materials exhibit $\chi_b < 0.05$. In contrast, for $\mathrm{SC}=\{\mathrm{GaAs},\,\mathrm{ZnSe}\}$ (see Suppl. Mater., Sec. F, Table S2 - S6), the relative ordering between the TMR values is preserved; however, when $\Delta_{xs}^{l} \leq \Delta_{xs}^{r}$, this inequality remains valid independently of whether the condition $E_F^{l} < E_F^{r}$ is satisfied (with the exception of the Co/SC/$\mathrm{Ni}_{80}\mathrm{Fe}_{20}$ case).

Similarly, when the ferromagnetic electrodes are fixed and the SC is varied, an ordering pattern emerges in the maximum TMR: the PSV with the largest TMR is typically the one with GaSb as the SC, followed by InSb (with TMR differences between these two ranging from 0.023\% to 0.37\%), and then InAs, GaAs, and ZnSe (see Suppl. Mater., Sec. F, Table S7 - S11). The exceptions correspond to the structures $\mathrm{Ni}_{80}\mathrm{Fe}_{20}$/SC/$\mathrm{Fe}_{90}\mathrm{Cr}_{10}$, Ni/SC/$\mathrm{Fe}_{90}\mathrm{Cr}_{10}$, and Ni/SC/Fe. Based on the analyses carried out in this work, it is not possible to identify a definitive cause for this ordering.

Finally, the agreement between the theoretical model proposed in this study for calculating the TMR in PSV structures and more complex related works supports the strong viability of the present approach. It also suggests that future theoretical improvements will allow a more detailed understanding of the interfacial band structure by incorporating relativistic SOC as the key element governing spin transport, and will enable more accurate predictions of the PSV response to variations in the SC thickness, the electron injection energy, and the magnetic configuration of the FM electrodes.
\section*{Acknowledgments}
The authors would like to thank the Master's Program in Physical Engineering of the Facultad de Ciencias \mbox{Naturales}, Exactas y de la Eduación of the Universidad del Cauca, Colombia, and the Instituto de Física La Plata (CONICET) Argentina, for providing the space and time for the development of this work.

\bibliographystyle{apsrev4-2}
\bibliography{Reference}

\end{document}